# Cross layer Interaction Models for SCTP and OLSR

M. Issoufou Tiado, R. Dhaou and A.-L. Beylot

*ENSEEIHT – IRIT, 2 rue Camichel BP 7122, 31071 Toulouse Cedex France +33 56158 8306*

[Issoufou_tiado@yahoo.fr](Issoufou_tiado@yahoo.fr), [Riadh.Dhaou@enseeiht.fr](Riadh.Dhaou@enseeiht.fr), [beylot@enseeiht.fr](beylot@enseeiht.fr)

**Abstract**

*The evolution from wired system to the wireless environment opens a set of challenge for the improvement of the wireless system performances because of many of their weakness compared to wired networks. To achieve this goal, cross layer techniques are used to facilitate the sharing of information between the layers of the OSI model. In some precedent works, the Reverse Cross Layer (RCL) method has been proposed to facilitate the design of cross layer conceptual models. The method has the advantage to highlight the impact of each cross layer interaction on each protocol in order to update its source code and to describe the intuitive gains that can be achieve. The method may be applied to a given protocol stack or to an existent cross layer model to integrate new interactions. In this paper, we are applying the RCL method on the stack that uses the Stream Control Transport Protocol (SCTP) at the transport layer and the Optimized Link State Routing (OLSR) at the network layer. Cross layer conceptual models are produced based on new cross layer interactions that are proposed to populate the environment subsystem built with the application of the RCL method. The improvement of the environment subsystem is specified through the performance gains provide by the new interactions. The implementation of the interactions that impact the SCTP protocol is described in the Interaction Description Array.*

*After the introduction, Section II of this paper presents an overview of the SCTP protocol. Section III is related to the overview of the OLSR protocol. Section IV is used for the application of the RCL method and the different interaction arrays it generates. Section V presents the improvement of the environment subsystem and the definition of the performance gain of each Cross Layer Atomic Action (CLAA).*

## 1. Introduction

The application of the RCL method specified in [1] and [2] allows to save the architecture benefit when building cross layer models. The layered architecture provides the modular conception frame, the systematic definition of the architecture components, the long terms objectives of using the network. The using of the RCL method allows to avoid the trap of an unbridled cross layer conception that will generate a "spaghetti" system [3] so that every modification for the improvement of the built system becomes difficult because of the possibility to generate many other modifications. Thus instead of modifying only one layer, all the system will be subjected to modifications. The cross layer conceptual models generate by the application of the RCL method allow to evaluate the complexity for the protocols stack to integrate the ensued cross layer interactions.

The first step of the RCL method allows to define the stack of protocols that must be used in the production of the cross layer interaction models. For the definition of new cross layer interactions to populate the proposed environment subsystem, the stack used in [1] is modified in this paper. SCTP is used at the transport layer, OLSR with the Internet Protocol (IP) [4] at the network layer, IEEE 802.11 for lower layers. To achieve the objective of an efficient presentation goal, all the RCL method steps are not developed. Only some models are produced to help to highlight the news interactions generate by the operation of SCTP at the transport layer associates to OLSR at the network layer. In remind, for the interaction census between the previous specify protocols, the classification provides in [1] and [2] will be used in order to distinguish the Activable Services CLAA (AS-CLAA), the Exported States CLAA (ES-CLAA) and the Notified Events CLAA (NE-CLAA). All these atomic action are identified to populate the proposed environment subsystem so that the stack performance improvement becomes highlighted.

In the following point, the overview of SCTP and OLSR is presented to facilitate the CLAA census. Some interesting points of the operation of the two protocols are discussed. Then the RCL cross layer arrays are produced before summarizing the points that improve of the environment subsystem.

For the contribution needs, this paper aims to highlight the global interaction to be installed between the protocols of the studied stack and to show the way by which the exploitation of each interaction by a particular protocol can intuitively improve the performance of the global system. The produced cross layer conceptual model initializes the work of evaluation to be conduct in order to confirm the intuitive gains.

## 2. Overview of SCTP

The SCTP protocol [5][6][7][8] is another variant of reliable transport protocol. The difference between SCTP and the Transmission Control Protocol (TCP) [9][10][11] comes from the based conception philosophy of the SCTP. The protocol is conceived for the wireless environment providing similar function than TCP, but using additional patterns such as association management and path lost management. SCTP uses the concept of stream and association. A stream is defined as a sequence of messages that must be transmitted in order. As a broader concept than a TCP connection, an association is a group in which each flow endpoint provides a list of transport addresses (@IP + port). Before transferring data, the SCTP sender and receiver execute sequences for the establishment of an association.

SCTP Operations
The SCTP protocol uses multiple function, namely, the transferred data control function, the error correction function, the congestion control mechanism, the priority management function, the congestion management function, the path management function.
The error management function of SCTP is based on the handling of the data retransmission triggered either by the expiry of the retransmission timer or by the reception of a Selective ACKnowledgement (SACK) indicating that the data has not been received. To reduce a potential congestion, the retransmission timer "Retransmission Time Out" (RTO) is adjusted on the base of the "Round Trip Delay" (RTD). In an active association with a fair and consistent transmission of data, the SACK messages generate more retransmissions than the timer expiration. To reduce the possibility of unnecessary retransmissions, the rule of four SACK is used, so that the retransmission is triggered upon the reception of the fourth SACK indicating the loss of data and avoiding retransmission for cases of reordering.
The path management function allows SCTP to establish the availability of a target node in the network. To detect such path lost, SCTP uses a count of unacknowledged number of retransmissions until the maximal attempt is reached before declaring the address inactive and notifying the information to the application. In such case and for multi-homed nodes, the protocol will use the alternative address if available to send data. The detection of the path lost is not only performed by the count of the number of retransmissions. Periodic messages called Heartbeat are regularly sent to all available destinations (for example toward the different addresses of the node). The SCTP protocol maintains update the number of sent Heartbeat messages to an unreachable destination without receiving acknowledgement. When that number reaches a predefine limit, the destination address is declared inactive. The Heartbeat messages continue to be sent to the inactive destinations until an ACK is received, and thus, the address is re-declared active. The rate of sending Heartbeat messages is associated to the RTO estimation to which is added a delay parameter that brings the Heartbeat traffic to be adapted to the needs of applications. The "Node unavailable NE-CLAA" is censed to populate the environment subsystem. This CLAA is for the application layer use because it is supposed that the Heartbeat messages initiate by SCTP are conveyed trough routing protocol primitives that will bring the routing level to detect the unavailability of the destination and to update its routing tables.

The Checksum Calculus of SCTP
The SCTP protocol uses the checksum calculus to certify that no modification of bits occurs during the transmission. The checksum is calculated so that it evenly and smoothly spreads its input packets over the available checked bits. Some weakness has been detected in [12] when using the Adler-32 SCTP checksum defined in [13]. The SCTP checksum calculus is updated and replaced to use a 32 bit CRC checksum [14]. Because with the Adler-32, particularly shorts packets, all packets having checksum values very close to 32640 make the calculus mechanism to fall under the higher likelihood of a small error to be left in a damaged packet with a valid checksum, than if all checksum values are equally likely. And as SCTP generates signaling messages usually less than 128 bytes, this weakness is exacerbated. The new CRC-32c replacing the current Adler-32 algorithm uses a polynomial division. The algorithm transforms an input message bit-string to a polynomial from which the CRC is calculated on-the-wire bit order using polynomial arithmetic [15]. As for the link layer the bit on the wire order is used and the first bit is associated to the high-order coefficient, for SCTP at the transport a convention is established for mapping SCTP transport messages to polynomials for the purposes of CRC computation. The 'mirrored' or 'reflected' [16] bit ordering is used: the first byte of the SCTP message provides the eight highest coefficients; within each byte B, the least-significant bit gives the most significant polynomial coefficient in B, and the most-significant bit is the least significant polynomial coefficient in B.
The "Common Checksum Calculus NE-CLAA" is conceived to make possible the study of the conditions under which the SCTP can share the CRC computation with the link layer.

The CLAA list of SCTP

The "Node unavailable Notified Events" is a new interaction produced by the path management function of SCTP. The interaction is added to populate the environment subsystem. Other CLAA used by TCP in the previous model in [1] and [2] are conserved with the same exploitation philosophy. The cross layer conceptual arrays produce later give the summary of all the interaction from or towards the transport layer. In addition to those CLAA, the OLSR protocol triggers some interactions used by SCTP that are presented below.

The "Common Checksum Calculus NE-CLAA" is also a new added CLAA, triggered by the link layer if the packet is certified to be out of error. The CLAA is destined to be used by SCTP. The point below relative to the improvement of the environment subsystem discusses about the conditions under which the Common Checksum Calculus can be conduct between the link layer and the SCTP protocol.

## 3. Overview of OLSR

The emission of periodic HELLO messages and the MultiPoint Relays (MPRs) concept are two main principles of the normal activities of the OLSR protocol [17]. The MPR concept is used to reduce the periodic messages transmission overhead. Because the MPRs are selected nodes designated to broadcast the periodic messages instead of performing the classical flooding mechanism in which every node retransmits each message when it receives the first copy. The elected MPR gives the link state information to all its partner nodes.

The periodic HELLO message broadcasting activity allows a node to perform the following three independent tasks: the link sensing, the neighbor detection and the MPR selection signaling. These tasks remaining on the periodic information exchange within a node's neighborhood serve to "local topology discovery". The Local Link Set (LLS), the Neighbor Set (NS) and the MPR Set are maintained by a node and used to generate HELLO message within HELLO_INTERVAL parameter smaller than or equal to REFRESH_INTERVALL. Many fields of the superstructures maintained by the OLSR are used to indicate the expiration or the validity time of the information that they are holding. In the LLS, the field "L_local_iface_addr" is the interface where the HELLO message is to be transmitted if the "L_time" field is greater than or equal to the current time meaning that the link is not expired. The expiration of both the "L_SYM_time" and the "L_ASYM_time" fields of the LS structure indicates that the link is lost. The expiration of the only "L_SYM_time" field brings the protocol to declare the link asymmetric. The neighborhood information base stores information about immediate neighbors, 2-hop neighbors, MPRs and MPR selectors. A record in the 2-hop Neighbor Set associate the main address of an immediate neighbor to the main address of a 2-hop neighbor, with in addition, the time at which the tuple expires and must be removed. The MPR Selector Set associates the main address of a node which has selected this node as MPR and the "MS_time" field that is given the time at which the tuple expires and must be removed. The Topology Set contains the main address of a node which may be reached in one hop from the other associated node and its "T_time" field specifies the validity time of the tuple before it is removed.

The expiration of one of the information in the OLSR superstructures can be used in the environment subsystem to trigger the "unavailable link NE-CLAA" in response to a message to send to that destination. The information maintains by the OLSR superstructures are shared between layers by the "superstructures ES-CLAA".

The "HTime" field in the HELLO packet specifies the time before the transmission of the next HELLO packet (HELLO emission interval) on a particular interface. The "VTime" field of the HELLO packet indicates how long time after the reception of the message the node must consider the information valid. These two fields of the HELLO packet are used to determine the condition under which the OLSR will trigger the "Common Signalization ES-CLAA" that aims to put into association the periodic broadcast of Heartbeat with the periodic emission of the HELLO packet. The point relative to the improvement of the environment subsystem discusses about the conditions under which the Common Signalization interaction can be used.

A model of the evaluation of the wireless link state is developed by the OLSR protocol in its hysteresis strategy. That strategy helps to take into account the instability nature of the wireless link. A case can occurred when from time to time, the link let HELLO messages pass through and fade out just after, and makes the neighbor information base to keep a bad link during the "validity time". To resolve that problem, the OLSR protocol computes the link quality through the "L_link_quality" field of the Link Tuples. This field is compared to the two following thresholds fixed between 0 and 1: HYST_THRESHOLD_HIGH, HYST_THRESHOLD_LOW with HYST_THRESHOLD_HIGH >= HYST_THRESHOLD_LOW. A link can be dropped based on either the expiration of the timer or based on the "L_link_quality" field dropping below HYST_THRESHOLD_LOW. The measure of the signal/noise ratio (SNR) is used if it is available to maintain and store the link quality in the "L_link_quality" field. An alternative to the SNR is to use the algorithm built with an exponentially smoothed moving average of the transmission success rate. The algorithm allows a node to use the stability rule upon receiving an OLSR packet and the instability rule when it is lost. The lost of OLSR packet is detected from the missing Packet Sequence Numbers on a per interface basis and by "long period of silence". The "long period of silence" derives from the absence of the reception of an OLSR packet during HELLO emission interval computed from the "Htime" field of the last received HELLO message.

The CLAA census of OLSR

The presentation of the OLSR protocol operations allows to determine the interaction that the protocol can generate to populate the environment subsystem. The studied stack has a real interest by the fact that the two protocols, namely the SCTP and the OLSR, use a periodic signalization to determine the availability of nodes in the network. That brings to an interesting question according to the use a common signalization when using cross layer mechanism between the two protocols, instead of the scheme by which every protocol generates its own additional traffic giving more overhead. To answer that question and as OLSR is a proactive routing protocol, it is possible for the protocol to export its superstructures and notified other interesting event toward other layers as for the SCTP use to avoid Heartbeat emission destined to a particular node under some conditions. One condition can be the notification of the availability of the node in the OLSR superstructure.

The following CLAA are censed initiate by OLSR: "Superstructures ES-CLAA", "Unavailable link NE-CLAA", "Wireless link status ES-CLAA" and "Common Signalization CLAA-ES". The "Superstructures ES-CLAA" allows OLSR to share its routing information with other layer through the environment subsystem. The "Unavailable link Notified Events" is derived from the expiration of one of the information in the OLSR superstructures. The "Wireless link status Exported States" comes from the hysteresis strategy by which the OLSR protocol evaluates the wireless link state and the "Common Signalization Exported States" is deriving from the HELLO messages emission and reception so that during a specified period and for the owner node of a HELLO packet, the SCTP suspends the emission of Heartbeat packet toward that node.

## 4. Application of the RCL method

### 4.1. CLAA census

In the previous studies realized in [1] and [2], the following NE-CLAA were proposed and are being kept here for the building of the new cross layer conceptual models with the new stack of protocols: the Explicit Congestion Notification (ECN) and Explicit Lost Notification (ELN), the jitter of sent packets, the retransmission avoidance (for saturation reasons, IP layer handoff, or other reasons that need retransmission and new traffic admission freezing), the link layer acknowledgement (used by OLSR and based on the link layer acknowledged frames containing complete IP datagram by using the Short Inter Frame Spacing-SIFS intervals of 802.11), the significant energy lowering event. The list of maintained CLAA is completed with the following CLAA used previously by TCP and kept here for the SCTP use with the same exploitation principle. The maintained CLAA are: the "Packet loss ratio ES", "SNR ES", "Bit Error Rate (BER) ES", "Received Signal Strength (RSS) ES", "Energy level ES".

In addition of the previous listed CLAA destined to populate the environment subsystem, the SCTP produces the "Node unavailable Notified Events" and uses the "Common Checksum Calculus NE-CLAA" of the link layer. The OLSR protocol adds the following CLAA: the "Superstructures Exported States", the "Unavailable link Notified Events", the "Wireless link status Exported States" and the "Common Signalization Exported States".

For the application of the RCL method as describe in [1] and [2], the steps following the CLAA census are summarized. The sections below present only the Protocol Interaction Array, the Function Interaction Array and the Interaction Description Array. The three cross layer conceptual models help to get a complete view of the interaction between layer and bring to the definition of the use of each interaction by the SCTP protocol. The intuitive idea is that for every interaction, a performance gain of the global system can be achieved and must be implemented and evaluated in further work initialize by this paper.

### 4.2. Protocol Interaction Array

The production of the Protocol Interaction Array takes into account the use of SCTP at the transport level instead of TCP. The CLAA presented below show the interaction between the protocols stack.

| Cross layer Atomic Action (CLAA) | Protocols | | | | | |
|---|---|---|---|---|---|---|
| | Application | SCTP | OLSR | IP | Link 802.11 | Physical 802.11 |
| Node unavailable Notified Events | D | S | | | | |
| Jitter of sent packets Notified Events | | D | | | S | |

|  | Protocols | | | | | |
|---|---|---|---|---|---|---|
| **Cross layer Atomic Action (CLAA)** | Application | SCTP | OLSR | IP | Link 802.11 | Physical 802.11 |
| Retransmission avoidance Notified Events |  | D | D |  | S |  |
| Acknowledgement Notified Events |  | D3 | D2, S3 |  | D1, S2 | S1 |
| Explicit Congestion Notified Events |  | D |  | S R. |  |  |
| Significant energy decrease Notified Events | D | D | D | D | D | D |
| Unavailable link Notified Events | D | D | S |  |  |  |
| Superstructures Exported States | U | U | S |  |  |  |
| Wireless link status Exported States | U | U | S |  |  |  |
| Common Signalization Exported States | U | U | S |  |  |  |
| Packet loss ratio Exported States | U | U |  |  | S |  |
| SNR Exported States | U | U |  |  | U | S |
| RSS Exported States | U | U |  |  | U | S |
| BER Exported States | U | U |  |  | U | S |
| Energy level Exported States | U | U | U | U | U | U |
| FEC Activable Service |  | U |  |  | S |  |
| Common Checksum Calculus Notified Events |  | D |  |  | S |  |
| ARQ Activable Service |  | U |  |  | S |  |

Legend : R. = Remote
**Table 1.** Cross-Layer Atomic Action Array

### 4.3. Function Interaction Array

The function interaction array of SCTP is produced according to the protocol's functions. It represents the conceptual model of interactions between the SCTP functions and the censed CLAA. To refer to the RCL method, that array allows to precise the SCTP function impacts by every CLAA.

|  | SCTP functions | | | | Other Protocols | | | | |
|---|---|---|---|---|---|---|---|---|---|
| **Cross layer Atomic Action (CLAA)** | Transf. Data Ctrl | Error Correct. | Congest Control | Path Manag | Application | OLSR | IP | Link 802.11 | Physical 802.11 |
| Node unavailable NE |  |  |  | S | D |  |  |  |  |
| Jitter of sent packets NE | D |  |  |  |  |  |  | S |  |
| Retransmission avoidance NE | D |  |  | D |  | D |  | S |  |
| Acknowledgement NE | D3 |  |  |  |  | D2, S3 |  | D1, S2 | S1 |
| Explicit Congestion NE |  |  | D |  |  |  | S R. |  |  |
| Significant energy decrease NE | D |  |  |  |  | D | D | D | S |
| Unavailable link NE | D |  |  | D |  | S |  |  |  |
| Superstructures ES | U |  |  | U |  | S |  |  |  |
| Wireless link status ES | U |  |  | U |  | S |  |  |  |
| Common Signalization ES |  |  |  | U |  | S |  |  |  |
| Packet loss ratio ES | U |  |  |  |  |  |  | S |  |
| SNR ES | U |  |  |  |  |  |  | U | S |
| RSS ES | U |  |  |  |  |  |  | U | S |
| BER ES | U |  |  |  |  |  |  | U | S |
| Energy level ES | U |  |  | U |  | U | U | U | S |

| Cross layer Atomic Action (CLAA) | SCTP functions | | | | Other Protocols | | | | |
|---|---|---|---|---|---|---|---|---|---|
| | Transf. Data Ctrl | Error Correct. | Congest Control | Path Manag | Application | OLSR | IP | Link 802.11 | Physical 802.11 |
| FEC AS | U | | | | | | | S | |
| Common Checksum Calculus NE | | D | | | | | | S | |
| ARQ AS | U | | | | | | | S | |

Legend : R. = Remote
**Table 2.** Function Interaction Array of SCTP

### 4.4. Interactions description arrays

The step 6 of the RCL method allows to deduce the Interaction Description Array of each protocol of the studied stack. The following array shows as the contribution of this paper, the exploitation of each CLAA by the SCTP protocol function to improve the performances of the global stack.

| CLAA | SCTP function | Exploitation of the CLAA by the SCTP function |
|---|---|---|
| Jitter of sent packets Notified Events | Transferred Data Control | Reset the waiting SACK timer, Freeze the transmissions and retransmissions for the new time, do not increment the retransmission counter for the same destination, do not emit Heartbeat packet for the new period, do not invoke the congestion control |
| Retransmission avoidance Notified Events | | Freeze the transmissions and retransmissions for the period specified in the message Reset retransmission timeouts without incrementing the counter |
| Node unavailable Notified Events | | Update the reachable/unreachable state of the node in the environment subsystem according to the value of the Heartbeat or retransmission counter |
| Acknowledgement Notified Events | | Anticipate the la transmission of new data if the destination node is directly accessible |
| Explicit Congestion Notified Events | Congest. Ctrl | Invoke the congestion control mechanism |
| Unavailable link Notified Events | Transferred Data Control and/or Path Management | Reset the waiting SACK timer, Freeze the transmissions and retransmissions for the new time, do not increment the retransmission counter for the same destination, do not emit Heartbeat packet for the new period, do not invoke the congestion control |
| Superstructures Exported States | | Consult the status of a node before transmitting/retransmitting data or Heartbeat |
| Wireless link status Exported States | | Consult the channel state before transmitting/retransmitting data or Heartbeat |
| Common Signalization Exported States | | Consult the HELLO fields before transmitting/retransmitting data or Heartbeat |
| Significant energy decrease Notified Events | | Modify the frequency of the retransmission and/or the rate of the transmission, Disable the use of Heartbeat messages emission mechanism. |
| Packet loss ratio Exported States | | Adjust the retransmission frequency and transmission output according to the high value of this parameter that is established by threshold (indicate channel state). Modify the rate of the Heartbeat message transmission |
| SNR Exported States | | |
| BER Exported States | | |
| RSS Exported States | | Use the link layer ACK if the threshold indicates that the destination is directly accessible |

| CLAA | SCTP function | Exploitation of the CLAA by the SCTP function |
|---|---|---|
| Energy level Exported States | | Modify the retransmission frequency and the transmission throughputs according to high value of this parameter that is established by threshold. Disable the Heartbeat emission |
| FEC Activable Service | | Cancel the data checksum calculus |
| Common Checksum Calculus NE | | |
| ARQ Activable Service | Error Correct. | Cancel the data error correction function if the destination is directly accessible. |

**Table 3.** Interaction Description Array of SCTP

## 5. The improvement of the environment subsystem

The initial environment subsystem produced in [1] and [2] is being improved in this paper with new censed interactions produced by SCTP and OLSR in order to produce some performance gains.

The "Node unavailable Notified Events" CLAA is produced by SCTP to populate the environment subsystem. By using that CLAA at the application layer, the objective is to improve the answer duration for the user benefit.

The following four CLAA are produced by OLSR: the "Superstructures Exported States CLAA", the "Unavailable link Notified Events", the "Wireless link status Exported States" and the "Common Signalization CLAA-ES". The "Superstructures Exported States CLAA" and the "Unavailable link Notified Events" allow an application or the SCTP protocol to take the knowledge of the status of a node so that it becomes possible to optimize the answer duration for the user benefit. The "Wireless link status Exported States" will be helpful for SCTP to regulate the data or the Heartbeat transmission and retransmission. By this, the energy lost by unsuccessful transmissions and retransmissions will be minimized. The "Common Signalization ES -CLAA" derived from the periodic reception of the HELLO packet by the OLSR protocol. During the time specify in the two fields "HTime" and "VTime" that specify respectively the time before the transmission of the next HELLO packet and the validity period of the HELLO packet, the SCTP must suspend the Heartbeat emission toward that destination. This will allow to optimize the unnecessary transmission of Heartbeat packet and thus will bring to save energy.

The Common Checksum Calculus between the link layer and the SCTP at the transport layer derives from the use of the CRC polynomial by the link layer in the arrival order and the use of the CRC polynomial mapped back into SCTP transport-level bytes. The CLAA initialize a starting point of conducting the study and the simulation that allow to evaluate the conditions under which the role of the PSCTP polynomial produces by SCTP can be ensured by the PLL polynomial of the link layer that takes into account the data and the PSCTP at the transport layer with in addition the bytes of the lower layers. The performance criterions will remain on the gain of time that can bring the system to inject more data in the same time.

## 6. Conclusion

The RCL method allows to conduct cross layer design in an efficient frame to avoid uncontrollable mixing of interaction. The conceptual interaction models produced by the method let appear the utilization generate by the interaction at different level of the OSI model stack. In this paper, we apply that method to a stack using SCTP at transport level and OLSR at the link layer. Different models of interaction have been product with new interactions that allow to improve the performance of the global stack. The improvement of the environment subsystem has been demonstrated with the definition of the gain attended when using each interaction.

For further work to be conduct, the current paper helps to initialize the simulation that can show the performance gains reachable when using each CLAA.